# Benchmarking of stainless steel cube neutron leakage in Research Center Rez

Michal Kostal*, Zdeněk Matěj†, Martin Schulc, Evžen Losa, Jan Šimon, Evžen Novák, František Cvachovec†, Vaclav Přenosil†, Filip Mravec†, Tomáš Czakoj, Vojtěch Rypar, Andrej Trkov*, Roberto Capote**

*Research Centre Rez, Hlavni 130, Husinec-Rez 250 68, Czech Republic, Michal.Kostal@cvrez.cz*
*† Masaryk University, Botanická 15, Brno 612 00, Czech Republic, Matej.Zdenek@mail.muni.cz*
*\* Jožef Stefan Institute, Jamova cesta 39, 1000 Ljubljana, Slovenia Andrej.Trkov@ijs.si*
*\*\*Nuclear Data Section, International Atomic Energy Agency, A-1400 Wien, Austria Roberto.CapoteNoy@iaea.org*

## I. ABSTRACT

The integral experiments covering the neutron leakage from geometrically simple assemblies with a $^{252}$Cf source inside are very valuable tools usable in validation of transport cross section data, since geometric uncertainties play a much smaller role in simple geometric assemblies than in complex assemblies as for example reactor pressure vessel geometry. Since $^{252}$Cf(s.f.) is standard neutron source, the uncertainties connected with the source neutron spectrum can be even neglected. The paper refers on validation efforts of neutron leakage from stainless steel block ~50×50×50 cm in Research Center Rez. Both the neutron leakage flux at a distance of 1 m from the center of the cubical assembly using stilbene spectrometry and the activation rates at different positions of the assembly were evaluated. In addition to experiments, main sources of uncertainty were identified and evaluated. The results of the stilbene measurements are consistent with the activation measurements results.

## II. INTRODUCTION

The paper describes benchmarking of neutron leakage flux from a stainless steel block with a $^{252}$Cf(s.f.) (spontaneous fission) neutron source placed in the center (see Fig. 1). The purpose of the study is to bring an additional set of experimental data to the historical measurements of pure iron in spherical geometries [1] used by nuclear data evaluators to test the consistency of their theories. Current benchmarking efforts and detailed results are summarized in the OECD-NEA report [2].

The stainless steel (AISI 321 type) was chosen because it is an important structural material in nuclear energy industry. Due to the high content of iron and chromium, such experiments can help to improve main components' nuclear cross-sections. The results of integral experiments in such simple geometries can be easily compared with calculations using different nuclear data libraries, making them very effective to test the validity of the neutron cross-section data. The reason is in fact that the integral quantity, such as the neutron flux or reaction rate, can usually be measured much more accurately than differential nuclear data, so it is tempting to use such data to refine the nuclide cross-section evaluations.

A cylindrical hole optimized for positioning the $^{252}$Cf(s.f.) neutron source is designed to position the neutron source into geometrical center of block. The neutron leakage was measured using a stilbene scintillator at the distance of ~ 1 m from the block center (the determined quantity was neutron flux density in the energy region 1 to 10 MeV), and by means of reaction rates on the surface of the block with the use of activation detectors using $^{115}$In(n,n') and $^{58}$Ni(n,p) reaction. The block is composed of individual plates which allow the measurement of spatial distribution of $^{58}$Ni(n,p) reaction rate in various deepness of the block.

## III. MATERIALS AND METHODS
### III.A. Stainless steel block
The experiments were carried out using the AISI 321 type stainless steel block (19.65 % Cr, 9.25 % Ni, 1.8 % Mn, 0.4 % Mo and 0.3 % Cu, Si, Ti each). The block is composed of plates with dimensions 50.2 × 50.2 cm and thicknesses of 3, 3.54, and 5.04 cm forming 50.4 cm thick assembly with 2.6 cm in diameter cylindrical hole with the length of 28.1 cm. These dimensions allow to centralize the $^{252}$Cf(s.f.) source with emission of 1.8E8 n/s in block, thus simplifying the transport model description. The view on the block with shielding cones is in Fig. 1. As the block is assembled from plates which are screwed together, totaling to the block with thickness of 50.4 cm,. The measurement with foils placed among stainless steel plates was carried out as well. The schematic assembly view with foils positions is plotted in Fig. 2.

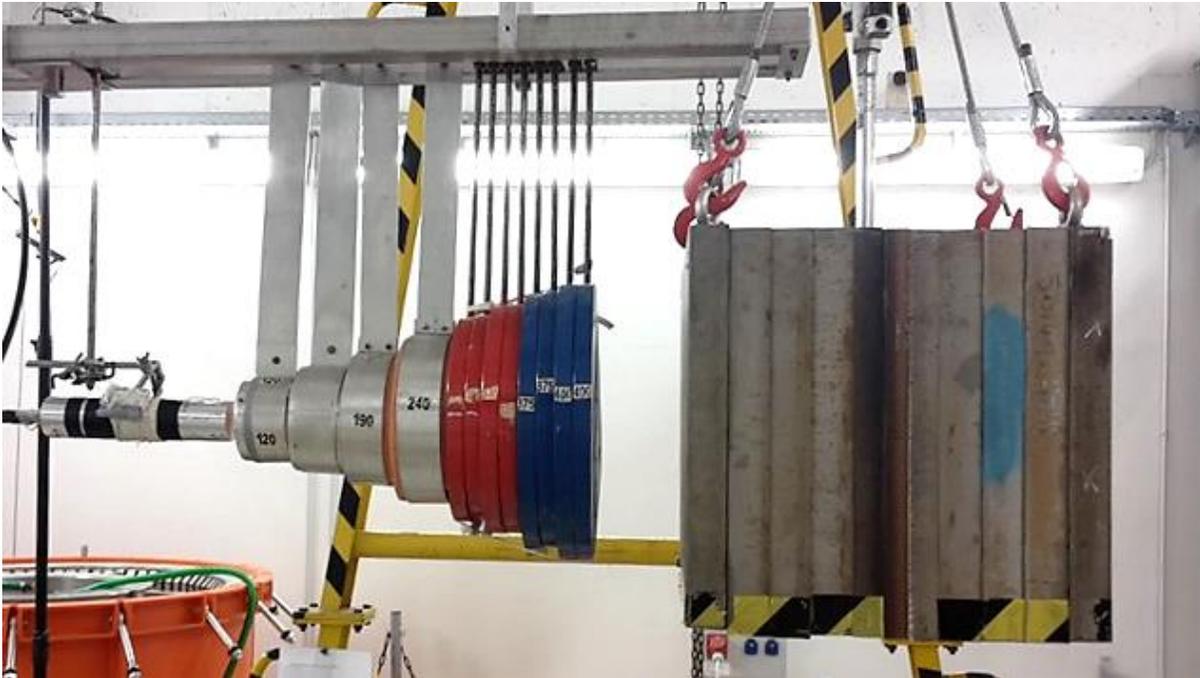
Fig. 1. Experimental configuration with stilbene detector, shadow cone and stainless steel block 50.4 × 50.2 × 50.2 cm

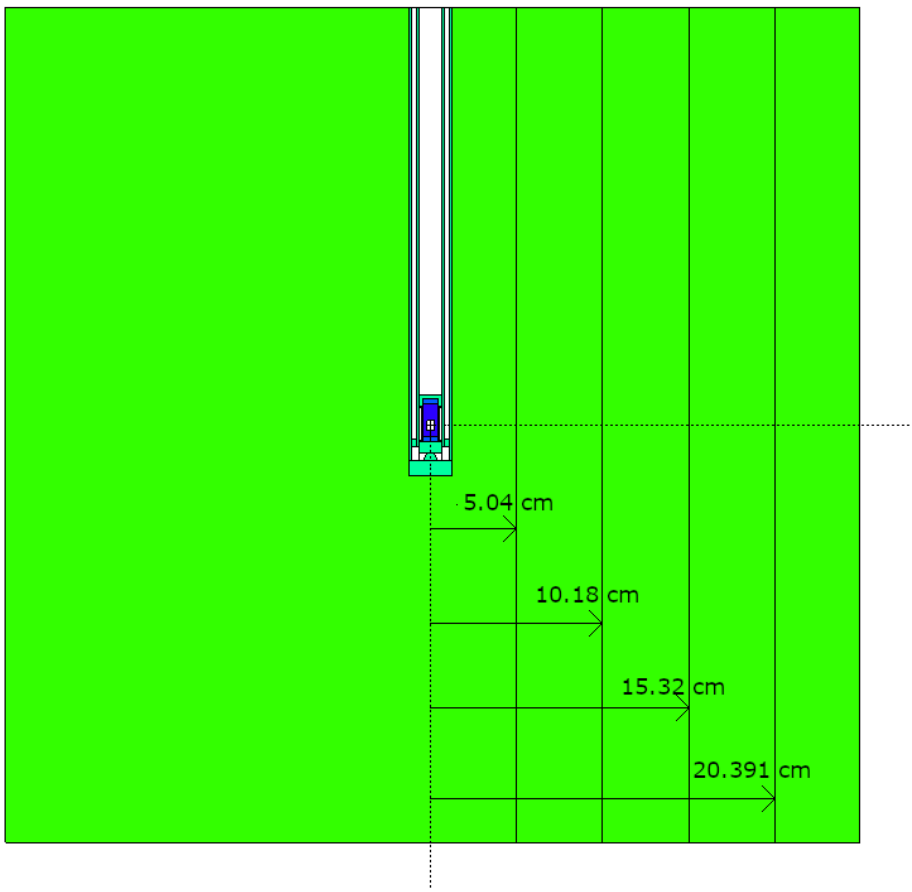
Fig. 2. Location of foils regarding to source position (X-Z cut), the source is in geometrical center

**III.B. Stilbene measurements**

The two-parameter spectrometric system NGA-01 [3] with stilbene cylindrical crystal (10×10 mm) was used in neutron flux density measurements. Neutrons are measured indirectly by means of recoiled protons, and photons by means of recoiled electrons. The Pulse Shape Discrimination (PSD) was used to distinguish the type of the detected particle, neutron or gamma, by analyzing the pulse shape (discrimination integral), whereas particle energy is evaluated from the integral of the total response. Due to the low gamma background, both signals are well separated [2]. Acquired recoiled proton spectra (or electron spectra in case of gamma field measurement) are then deconvoluted the Maximum Likelihood Estimation [4]. Apparatus for neutron spectrometry with the stilbene detector can be calibrated by gamma sources thanks to the known relationship between responses to neutrons and gammas. The energy calibration during measurement was done indirectly using a 2.23 MeV photon peak from $^1$H(n,γ) and a 0.66 MeV peak from a $^{137}$Cs source, and the rest of the peaks from $^{10}$B(n,α) (0.478 MeV) and $^{60}$Co (1.17, 1.33 MeV) are used for verification. The full calibration curve is plotted in Fig. 3. The way of the energy calibration was tested in the well-defined silicon filtered neutron beam in the LVR-15 reactor. A comparison of measured neutron spectrum with the reference neutron spectrum as established in past in LVR-15 reactor [5] is plotted in Fig. 4. The efficiency calibration was tested using Cf point neutron source [6]. The same source was used for estimation of deconvolution uncertainty, which was estimated being lower than 5 % [7].

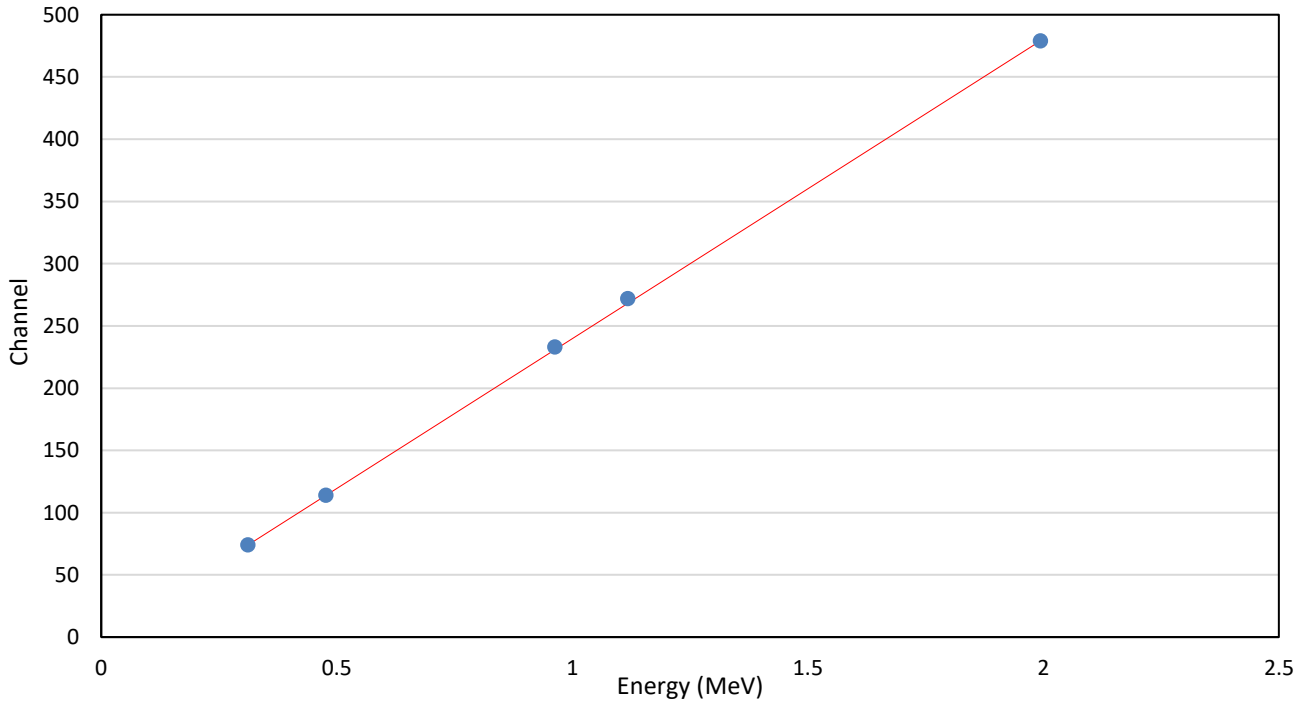

Fig. 3. Calibration curve in stainless steel experiment ($^{10}$B capture (n,α), $^{137}$Cs, $^{60}$Co, $^1$H(n,γ) )

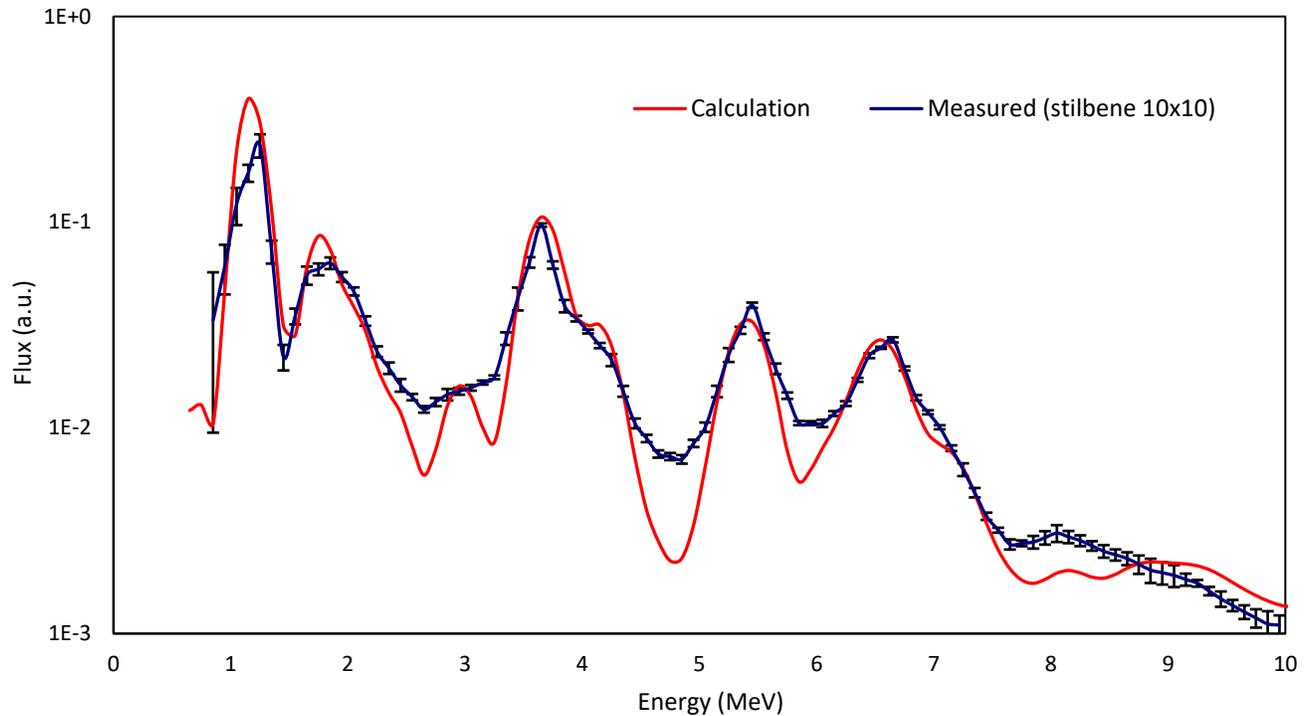

Fig. 4. Validation of deconvolution method in silicon filtered beam [7] and crystal 10 mm in diameter, 10 mm high.

The measured spectrum is affected by the presence of structural components (concrete walls and floor). To obtain a pure leakage spectrum, the room return effect was subtracted from the primary spectrum using shadow cones (see Fig.5). The cones are optimized to attenuate the primary neutrons by factor 500 – 1000, thus the transmission contribution to the background effect is negligible.

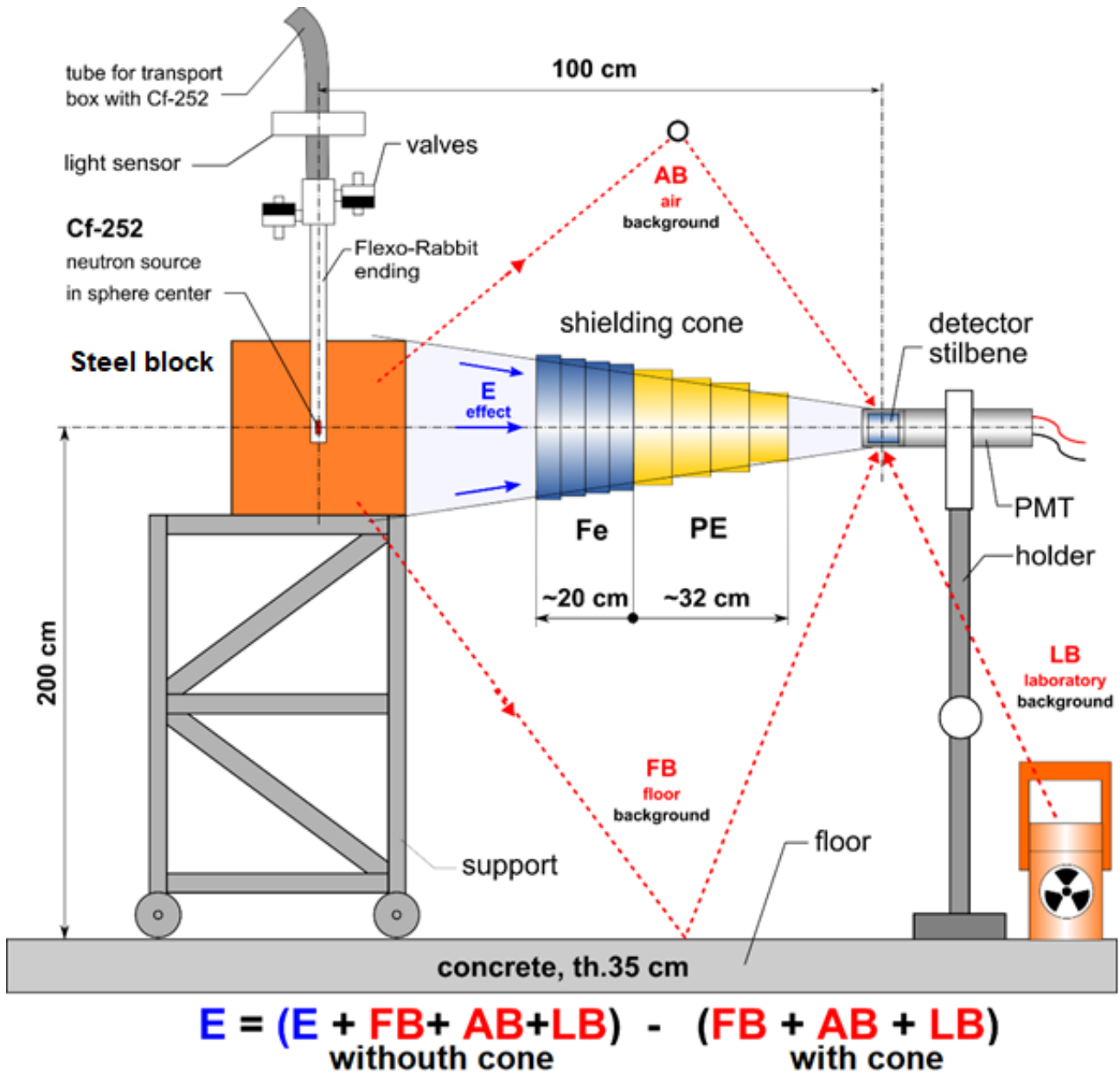

Fig. 5. Shadow cones for determination of room return fraction (X-Z cut)

The background (i.e., recoiled proton flux measured with cones) was subtracted directly from the total proton flux. The net proton flux with the extracted room effect was then deconvoluted. The experimentally determined and calculated room effect together with calculated transmission by cones are plotted in Fig. 6. The calculation was carried out using MCNP6.2 and ENDF/B-VIII.0. The background induced by shielding cone was found being negligible [2].

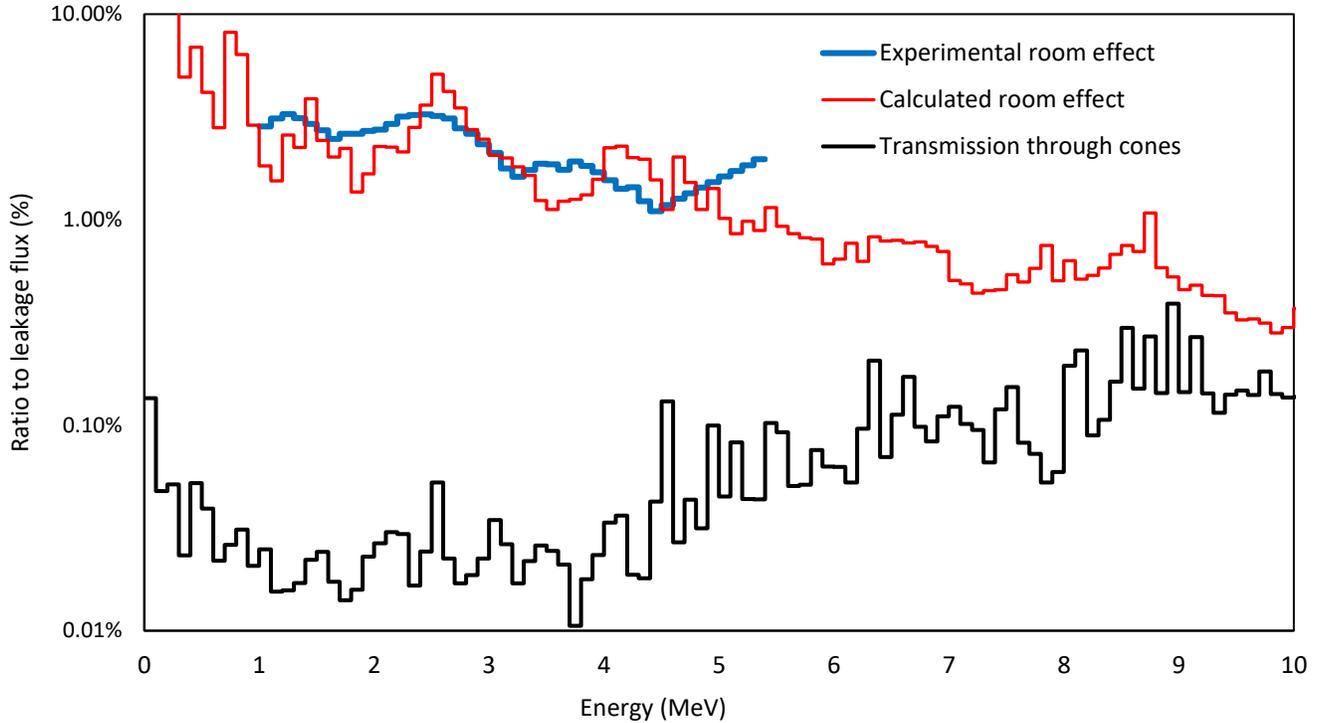

Fig. 6. Evaluated room effect and cones transmission

**III.C. Activation measurement**

The neutron leakage was monitored using reaction rates derived from activation foil measurement for validation of the stilbene results. Two reactions were used: $^{115}$In(n,n') and $^{58}$Ni(n,p). Nickel foil was also placed between plates in different thicknesses of stainless steel (see Fig. 6). The reaction rates were derived from the Net Peak Areas (NPA) of the irradiated activation foils and were measured in the end cap geometry, in which the measured sample is on the cap of the HPGe detector [2].

The resulting reaction rates on the block surface are listed in Tab. I, reaction rates in the inside positions of the block in Tab.II (see Fig. 2 with plotted activation foils positions in the block), their comparison with calculation using various nuclear data libraries in Tab. III and Tab. IV.

TABLE I. Reaction rates on cube surface

| Reaction | Mean (1/atom/s) | Rel. Unc. (%) |
|---|---|---|
| $^{115}$In(n,n') | 4.4E-30 | 3.41 |
| $^{58}$Ni(n,p) | 7.77E-31 | 5.09 |

TABLE II. Reaction rates of $^{58}$Ni(n,p) in inside positions of block

| Steel thickness | Mean (1/atom/s) | Rel. Unc. (%) |
|---|---|---|
| 5.04 cm | 2.28E-28 | 7.1 |
| 10.08 cm | 4.22E-29 | 3.3 |
| 15.12 cm | 9.77E-30 | 4.0 |
| 20.16 cm | 2.80E-30 | 3.7 |

**III.D. Calculations**

The calculations were carried out using MCNP6.2 [8] Monte Carlo code in combination with various nuclear data libraries ENDF/B-VIII.0 [9], JEFF-3.3 [10], JENDL-4 [11], and INDEN [12] for neutron transport through the material of the stainless steel assembly. Since INDEN library for stainless steel materials contains only cross-sections for chromium and iron, ENDF/B-VIII.0 has been used for the remaining materials. For the simulation of the particle transport from the source to the detector, variance reduction technique based on point detectors was used. 1E11 particle histories have been used for calculations resulting in good statistics in higher energy tail of the spectrum (less than 3 % in group 19.9 -20 MeV). The calculational model considers all details concerning the geometry of the source structural components. The neutron spectrum of the $^{252}$Cf(s.f.) source as well as dosimetry cross sections of the In and Ni monitors were taken from IRDFF-II [13] library. Estimation of uncertainties going out of the geometry and composition was done computationally by direct perturbations of the given parameters and ENDF/B-VIII.0 library has been used in that case.

## IV. RESULTS

The evaluated neutron spectra (see Fig. 7) were compared with the experiment using C/E-1 comparison (see Fig. 8). The plotted experimental uncertainty is covering uncertainty in calibration, deconvolution, and source emission [2]. The C/E-1 comparison of reaction rates on stainless block surface is listed in Tab. III, comparison in inner positions in Tab. IV. The calculated evolution of neutron spectra is plotted in Fig. 9. Based on its sensitivity in low energy region the indium dosimeter with $^{115}$In(n,n') reaction was used, thus supplementing the $^{58}$Ni(n,p) reaction for monitoring of the fast neutron flux density. The sensitivities to both reactions in cube surface are plotted in Fig. 10.

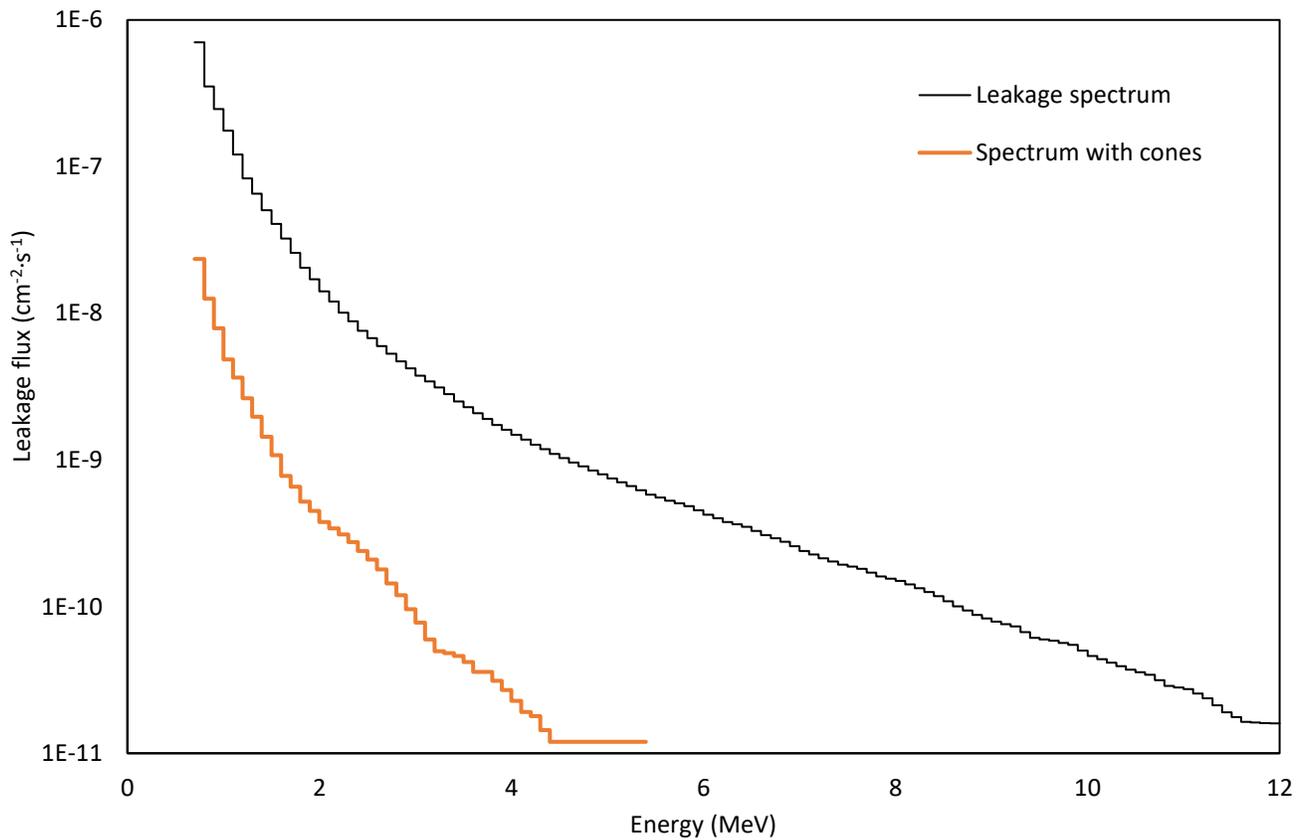

Fig. 7 Measured neutron leakage spectra

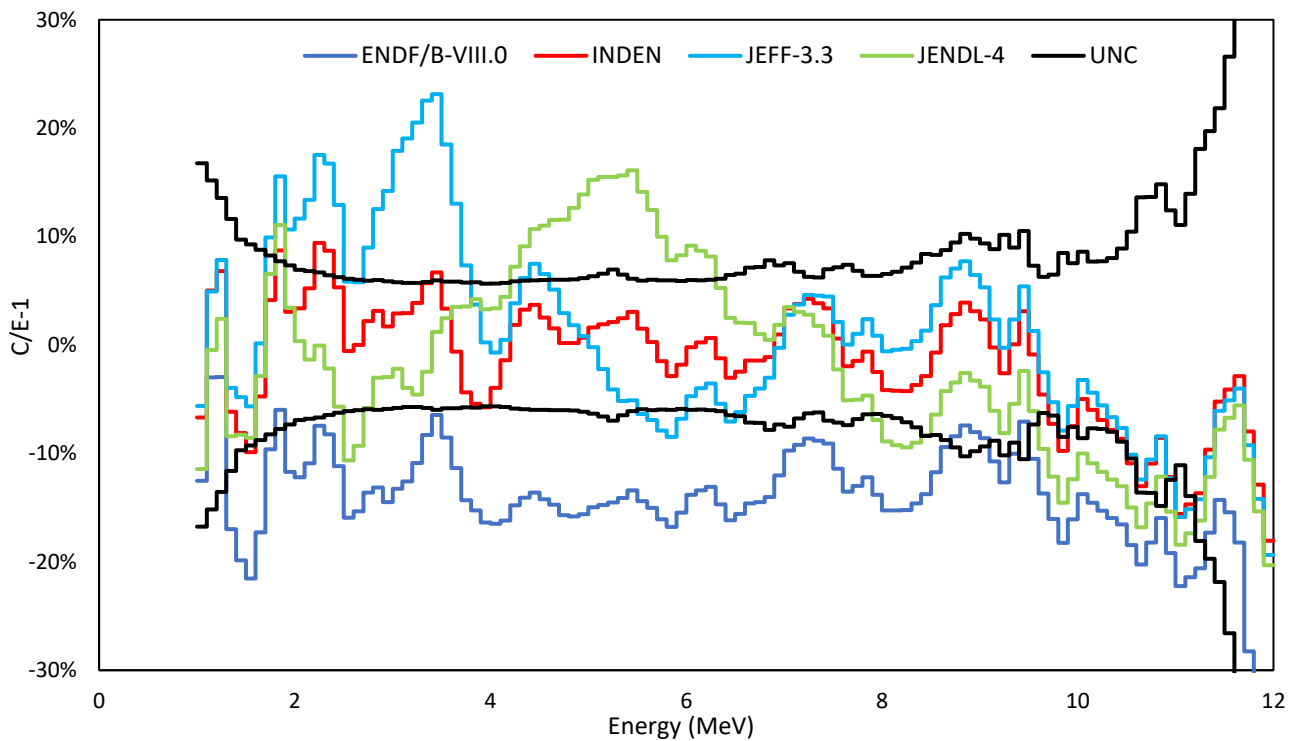

Fig. 8 Calculated and measured neutron leakage spectra by means of C/E-1 comparison

TABLE III. C/E-1 for foils on block surface in various libraries

|  | ENDF/B-VIII.0 | INDEN | JEFF-3.3 | JENDL-4 |
|---|---|---|---|---|
| $^{115}$In(n,n') | -15.5% | -5.7% | -1.5% | -9.0% |
| $^{58}$Ni(n,p) | -17.4% | -4.6% | 3.1% | -4.6% |

TABLE IV. C/E-1 for nickel foils with $^{58}$Ni(n,p) reaction in various libraries inside steel block

| Steel thickness | ENDF/B-VIII.0 | INDEN | JEFF-3.3 | JENDL-4.0 |
|---|---|---|---|---|
| 5.04 cm | -8.6% | -5.4% | -3.4% | -6% |
| 10.08 cm | -8.4% | -2.3% | 1.3% | -3% |
| 15.12 cm | -9.1% | 0.4% | 5.5% | -0.4% |
| 20.16 cm | -12.9% | -0.1% | 5.3% | -1.2% |

The best agreement in stilbene measurement is observed in the case of the INDEN evaluation. This result is consistent in both stilbene measurement and activation measurement. On the other hand, significant underprediction can be observed in ENDF/B-VIII.0. In the inside block positions, an increasing rate of underprediction with increasing distance from the neutron source can be observed in the case of the ENDF/B-VIII.0. In the case of activation measurement best agreement gives JEFF-3.3 library. The good agreement of reaction rates in INDEN evaluation can be observed as well. In general, except ENDF/B-VIII.0, no trend of agreement in dependence on material thickness is observable in the case of the other libraries.

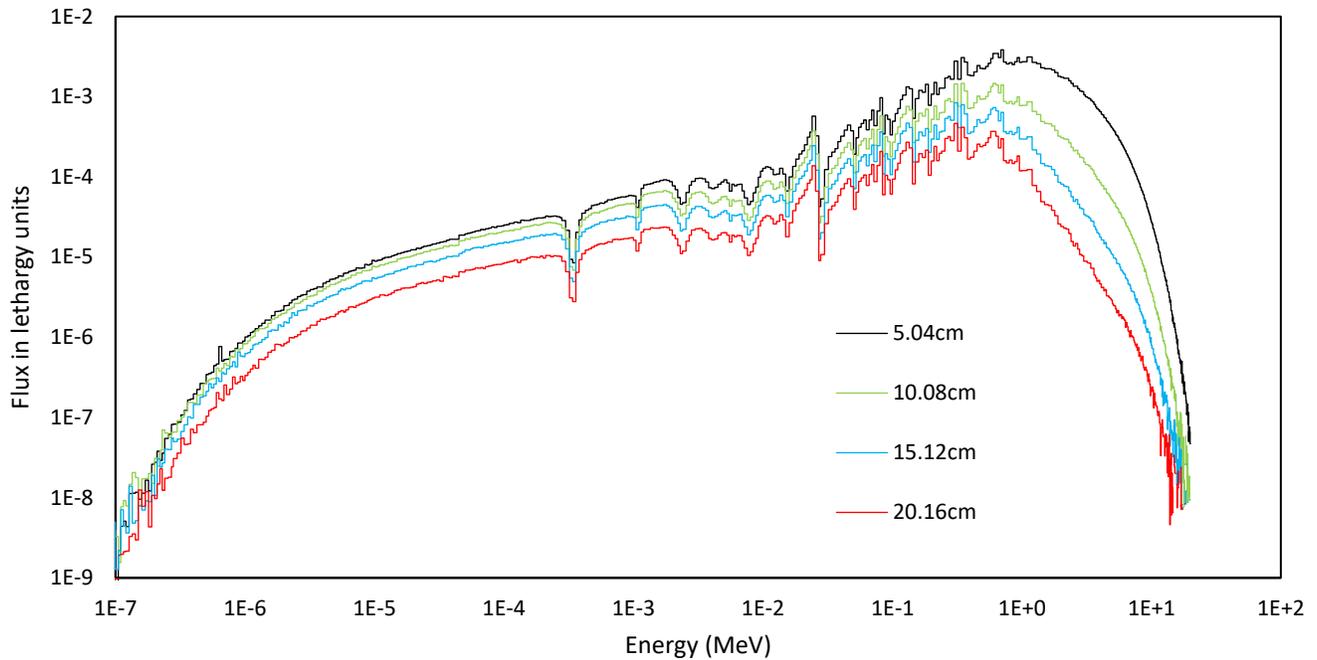

Fig. 9 Calculated neutron spectra in various deepness of steel using ENDF/B-VIII.0 library

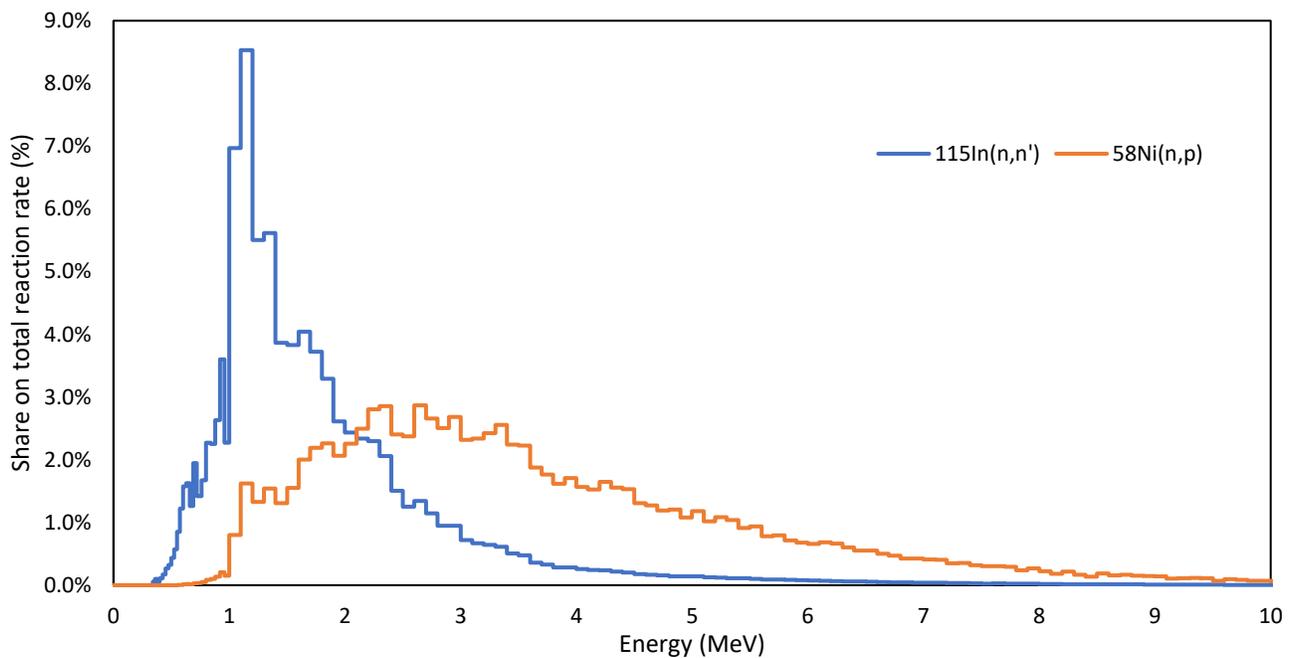

Fig. 10 Calculated sensitivity of used dosimeters on steel block surface using ENDF/B-VIII.0 library

## IV.A. Reliability of results

The experimental uncertainties plotted in Fig. 8 do not cover the parameter uncertainties. These parameter uncertainties are going out especially of the material and geometrical uncertainties of the steel block description and position of the detector during measurement. They are determined from material characterization methods as He pycnometry and XRF, or by engineering judgement [2]. Main contributing uncertainties are summarized in Tab. V. The coordinate system, used in description of uncertainties in Table V, is indicated in Figures 2 and 5.

TABLE V. Summary of related uncertainties

|  | Mean value | 1σ uncertainty | unit |
|---|---|---|---|
| Density (ρ) | 7.9083 | 0.0093 | g/cm³ |
| Thickness (X) | 50.4 | 0.2 | cm |
| Width (Y) | 50.2 | 0.8 | cm |
| Height (Z) | 50.2 | 0.8 | cm |
| Det. distance | 1000 | 0.5 | mm |
| Det. misalignment | 4 | 0.5 | mm |
| Fe wt. % | 0.6783 | 0.0052 | wt. % |
| Cr wt. % | 0.1965 | 0.0029 | wt. % |
| Ni wt. % | 0.0925 | 0.0024 | wt. % |
| Mn wt. % | 0.018 | 0.0021 | wt. % |

The effect of parameter uncertainty on the neutron flux at the stilbene detector position is determined by calculation using direct perturbation. The results for geometrical uncertainties are plotted in Fig.11, and the effect of material uncertainties is plotted in Fig 12. The effect of detector position uncertainty is not plotted, as the effect is below 0.1 %. The effect of experiment setup components on neutron flux is plotted in Fig. 13. The graph shows the effect of neglecting selected components in the calculational model on the leakage flux. Namely the effect of small gaps between the plates was studied, effect of minor components of stainless steel, and effect of source structural components like source stainless steel cladding and flexo rabbit components.

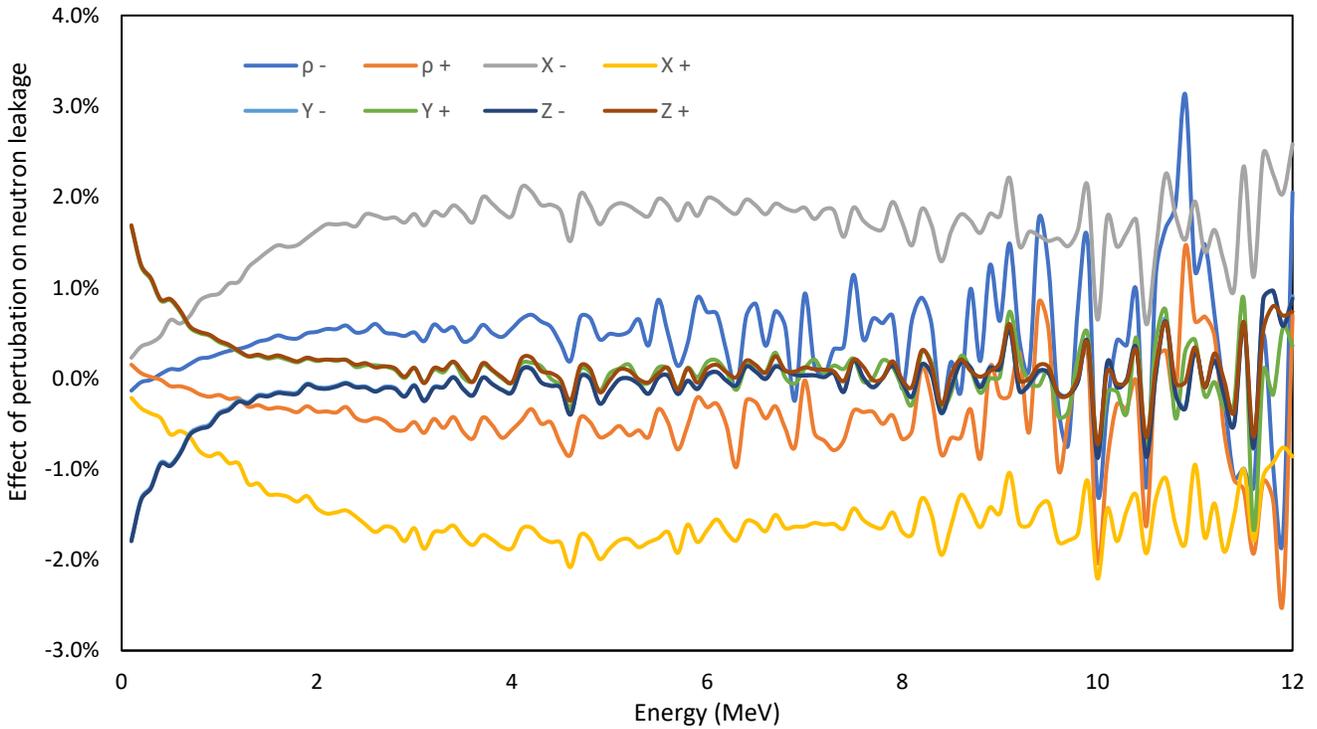

Fig. 11. Effect of geometrical uncertainties on neutron leakage

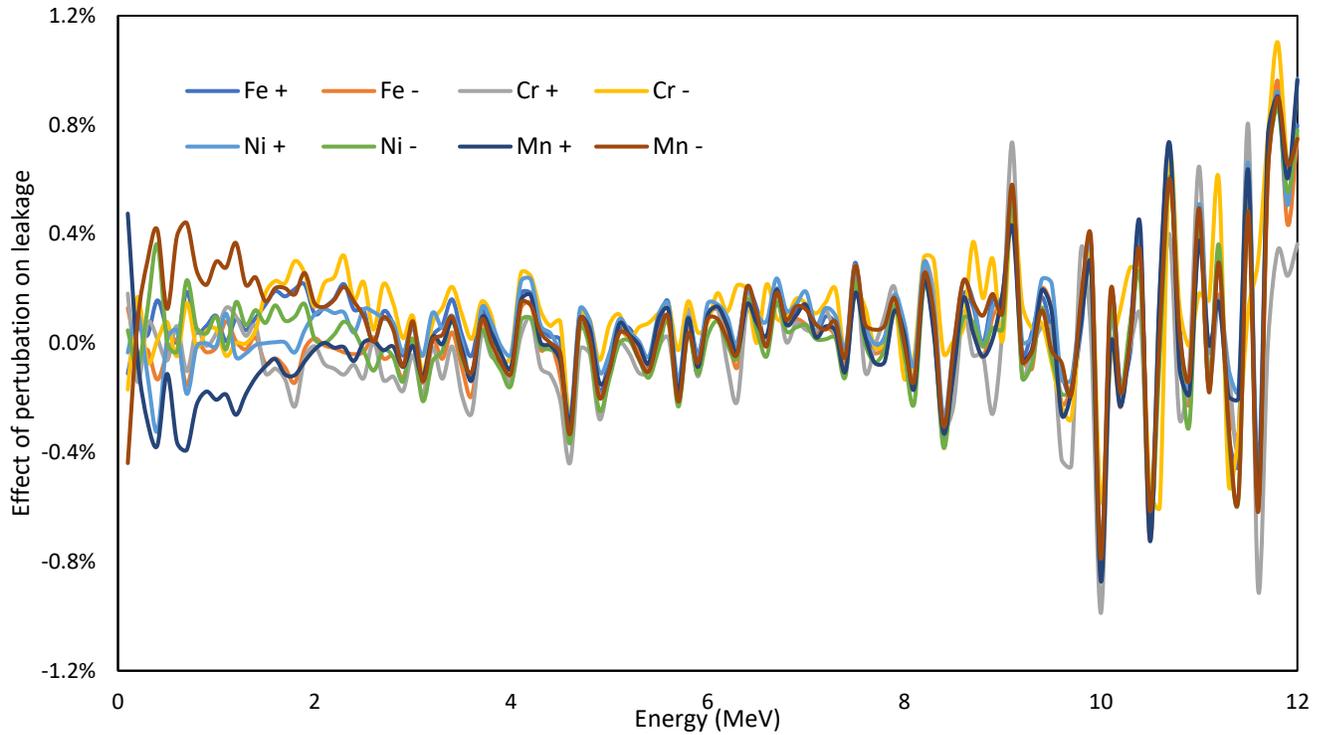

Fig. 12. Effect of material uncertainties on neutron leakage

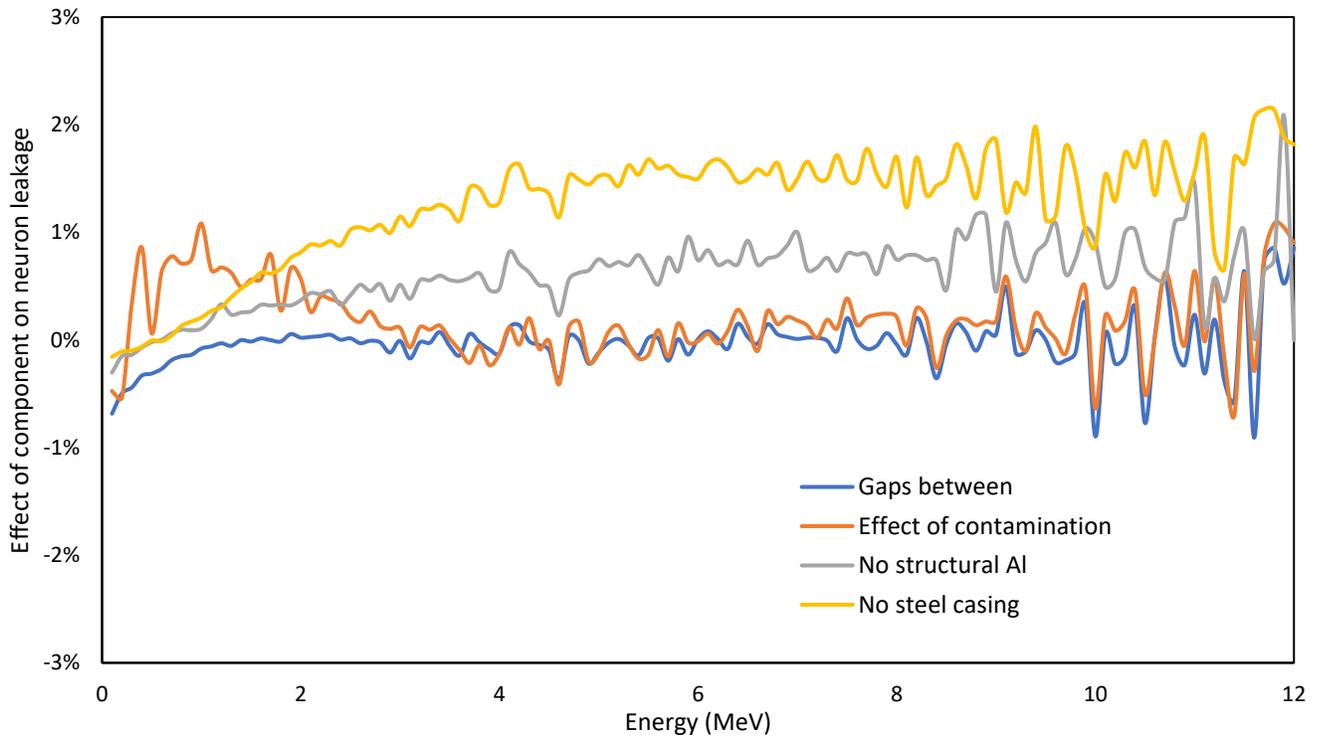

Fig. 13. Effect of components on neutron leakage flux

According to expectations, the most significant effect has uncertainty of the cube thickness although it is not higher than 2 %. The effect of uncertainties in material composition has a negligible effect on neutron leakage flux.

Structural components have a notable effect on the neutron flux. This means that it is important to include them in the calculational model. However, the effect is not higher than 5 % in the case of the effect of stainless steel casing of $^{252}$Cf(s.f.) source, so when occurs the uncertainties that change the abundance of materials of the source components not more than 5 %, the effect the leakage flux is not higher than 0.3 %.

In general, it can be concluded, that the parameter uncertainties have a significantly smaller effect on neutron flux than experimental uncertainties connected with stilbene measurement.

The effect of parameter uncertainty on the reaction rates in activation detectors is determined by calculation using direct perturbation, similar to the stilbene case. The results for the most significant uncertainty sources for foils on the surface are listed in Tab. VI, the results for foils in inside positions in Tab. VII. In addition to previous sources the uncertainty in the Y, and Z positions, which is 3 mm, plays a role in this measurement.

Similar to the stilbene case, the most significant effect has uncertainty of the cube thickness which effect is about 2 %. In the case of reaction rates inside the steel block, the most significant effect plays the uncertainty of the foil position regarding to the neutron source. This effect is most significant for the closest foils, where the uncertainty can be as high as 5 %. For more distant foils the parameter uncertainties do not exceed 1 %.

TABLE VI. Effect of parameter uncertainty on reaction rates for foils on block surface (%)

|  | Steel thickness | Gaps thickness | Steel thickness |
|---|---|---|---|
| $^{58}$Ni(n,p) | 0.4 | 0.9 | 2.3 |
| $^{115}$In(n,n') | 0.3 | 0.8 | 1.7 |

TABLE VII. Effect of parameter uncertainty on reaction rates in inside block positions (%)

| Steel thickness | Y pos. | Z pos. | ρ | Gaps thickness | Steel thickness |
|---|---|---|---|---|---|
| 5.04 cm | 3.9 | 5.3 | 0.4 | 0.2 | 0.1 |
| 10.08 cm | 0.5 | 0.4 | 0.2 | 0.5 | 0.1 |
| 15.12 cm | 0.5 | 0.4 | 0.5 | 0.4 | 0.1 |
| 20.16 cm | 0.2 | 0.3 | 0.5 | 0.5 | 0.4 |

**V. CONCLUSIONS**

The neutron leakage flux from the stainless steel cube with $^{252}$Cf(s.f) source inside was characterized by both stilbene and activation foil measurement.

The simulations using INDEN evaluation and JEFF-3.3 are in satisfactory agreement with the experiment, while both stilbene and activation measurements are in good agreement with each other. In case of INDEN this experiment was used for its validation.

The most significant source of uncertainty is experimental uncertainty covering uncertainty in calibration, deconvolution, and source emission. The parameter uncertainty has a relatively low effect which can be neglected in the case of stilbene measurement. This source of uncertainty mainly affects namely the activation measurement so it must be considered in this case.

**VI. ACKNOWLEDGEMENTS**

Presented results were obtained with the use CICRR project LM2023041 and by the SANDA project funded under H2020-EURATOM-1.1 contract 847552.